\input harvmac
\input psfig

\def \O {{\cal O}}

\def \N {{\cal N}}

\def \lr { \lref}

\def\np {  {\it Nucl. Phys.} }
\def \pl { {\it  Phys. Lett.} }

\def \pr  { {\it Phys. Rev.} }

\lr\KW{I.R. Klebanov and E. Witten, ``Superconformal Field Theory on
Threebranes at a Calabi-Yau Singularity,'' 
{\it Nucl. Phys.} {\bf B536} (1998) 199, {\tt hep-th/9807080}.}

\lr \KS  {S.~Kachru and E.~Silverstein, ``4d Conformal Field Theories
and Strings on Orbifolds,''  {\it Phys. Rev. Lett. } {\bf 80} (1998)
4855,
{{\tt hep-th/9802183}}.}
\lr\LNV {A.~Lawrence, N.~Nekrasov and C.~Vafa, ``On Conformal Field
Theories in Four Dimensions,'' {\it Nucl. Phys.} {\bf B533} (1998) 199,
{{\tt hep-th/9803015}}.}

\lr\kleb{
S.S. Gubser and I.R. Klebanov, ``Absorption by Branes and Schwinger
Theory,'' {\it Phys. Lett.} {\bf B413} (1997) 41,
  {{\tt hep-th/9708005}}; for a review see I.R. Klebanov,
``From Threebranes to Large N Gauge Theories,''
{{\tt hep-th/9901018}}. }

\lr\GKtwo{
S.S. Gubser and I.R. Klebanov, ``Baryons and Domain Walls in an
{\cal N}=1 Superconformal Gauge Theory,'' \pr {\bf D58} 
(1998) 125025, hep-th/9808075.}

\lr\BKL{V. Balasubramanian, P. Kraus and A. Lawrence,
``Bulk vs. Boundary Dynamics in Anti-de Sitter Spacetime,''
{{\tt hep-th/9805171}}.}

\lr\Vij{V. Balasubramanian, P. Kraus, A. Lawrence and
S. Trivedi, ``Holographic Probes of Anti-de Sitter Spacetimes,''
{{\tt hep-th/9808017}}.}

\lr  \jthroat{
J.~Maldacena, ``The Large N Limit of Superconformal Field Theories and
  Supergravity,'' {\it Adv. Theor. Math. Phys.} {\bf 2} (1998) 231,
{{\tt hep-th/9711200}}.}

\lr\Aha{O. Aharony, Y. Oz and Z. Yin, ``M theory on
$AdS_p\times S^{11-p}$ and Superconformal Field Theories,''
{{\tt hep-th/9803051}}; S. Minwalla, ``Particles on $AdS_{4/7}$
and primary operators in $M_{2/5}$-brane world volumes,''
{{\tt hep-th/9803053}}. }

\lr  \GKP{
S.S. Gubser, I.R. Klebanov, and A.M. Polyakov, ``Gauge Theory Correlators
  from Noncritical String Theory,'' {\it Phys. Lett.} {\bf B428} (1998)
105, {{\tt hep-th/9802109}}.}

\lr  \EW{
E.~Witten, ``Anti-de Sitter space and Holography,''
 {\it Adv. Theor. Math. Phys.} {\bf 2} (1998) 253,
 {{\tt hep-th/9802150}}.}

\lr\MP{D.R. Morrison and M.R. Plesser,
``Non-Spherical Horizons, I,'' hep-th/9810201.}

\lr\FMMR{D.Z. Freedman, S.D. Mathur, A. Matusis and L. Rastelli,
``Correlation Functions in the AdS/CFT Correspondence,''
{{\tt hep-th/9804058}}.}

\lr\FGPW{D.Z. Freedman, S.S. Gubser, K. Pilch and N. Warner,
``Renormalization Group Flows from Holography, Supersymmetry,
and a $c$-Theorem,''
{{\tt hep-th/9804017}}.}

\lr\Myers{R.C. Myers, ``Stress Tensors and Casimir Energies in the
AdS/CFT Correspondence,'' hep-th/9903203.}

\lr\BK{V. Balasubramanian and P. Kraus,
``A Stress Tensor for Anti-de Sitter Gravity,'' hep-th/9902121.}

\lr\DM{
M.R.~Douglas and G.~Moore, ``D-branes, Quivers, and ALE Instantons,''
{{\tt hep-th/9603167}}.}

\lr\Kehag{
A. Kehagias, ``New Type IIB Vacua and Their F-Theory Interpretation,''
{{\tt hep-th/9805131}}.
}

\lr\FF{B.S. Acharya, J.M. Figueroa-O'Farrill, C.M. Hull, B. Spence,
``Branes at Conical Singularities and Holography,''
hep-th/9808014. }

\lr\afm{
O.~Aharony, A.~Fayyazuddin and J.~Maldacena,
``The Large N limit of ${\cal N}=2$, ${\cal N}=1$
Field Theories from Threebranes
in F Theory,''
{{\tt hep-th/9806159}}.}

\lr\SG{S.S. Gubser, ``Einstein Manifolds and Conformal Field Theories,''
\pr {\bf D59} (1999) 025006, hep-th/9807164. 
}

\lr\RD{D. Jatkar and S. Randjbar-Daemi,
``Type IIB string theory on $AdS_5 \times T^{nn'}$,''
hep-th/9904187.}

\lr\Kim{H.J. Kim, L.J. Romans and P. van Nieuwenhuizen,
``The Mass Spectrum of Chiral $N=2$ $D=10$ Supergravity on $S^5$,''
\pr {\bf D32} (1985) 389.}

\lr\GT{
G. 't~Hooft, ``A Planar Diagram Theory for Strong Interactions,'' 
{\it Nucl. Phys.} {\bf B72} (1974) 461.}

\lr\AP{A.M. Polyakov, ``String Theory and Quark Confinement,''
{\it Nucl. Phys. B (Proc. Suppl.)} {\bf 68} (1998) 1, {{\tt
hep-th/9711002}}. }

\lr\Lee{S. Lee, S. Minwalla, M. Rangamani and N. Seiberg,
``Three-Point Functions of Chiral Operators in $D=4$,
${\cal N}=4$ SYM at Large $N$,''
hep-th/9806074.}

\lr\BF{P. Breitenlohner and D.Z. Freedman,
``Stability in Gauged Extended Supergravity'',
{\it Ann. Phys.} {\bf 144} (1982) 249.}

\lr\KH{I.R. Klebanov, ``Touching Random Surfaces and Liouville Theory,''
\pr {\bf D51}, 1836 (1995), hep-th/9407167;
I.R. Klebanov and A. Hashimoto, ``Non-perturbative Solution
of Matrix Models Modified by Trace Squared Terms,''
\np {\bf B434}, 264 (1995), hep-th/9409064.}

\lr\DFR{E. D'Hoker, D. Z. Freedman and L. Rastelli, 
``AdS/CFT 4-point functions: How to Succeed at z-integrals Without Really
Trying,'' hep-th/9905049.}

\lr\MW{W. M\" uck and K.S. Viswanathan, \pr {\bf D58}, 041901 (1998),
hep-th/9804035.
}

\lr\cd{
P.~Candelas and X.~de la Ossa, ``Comments on Conifolds,''
\np {\bf B342} (1990) 246.}

\lr\Ur{A. Uranga, ``Brane Configurations for Branes at Conifolds,''
hep-th/9811004; K. Dasgupta and S. Mukhi, ``Brane Constructions,
Conifolds and M-Theory,'' hep-th/9811139.}

\lr\GMS{
B. Greene, D. Morrison, A. Strominger, ``Black Hole Condensation and the
Unification of String Vacua,''
\np {\bf B451} (1995) 109.}

\lr\Romans{
L.~Romans, ``New Compactifications of Chiral $N=2$, $d=10$
Supergravity,'' \pl {\bf B153} (1985) 392.}

\lr\Dobrev{V.K. Dobrev, ``Intertwining Operator Realization of
the AdS/CFT Correspondence,'' hep-th/9812194.}

\lr\klt{P. Kraus, F. Larsen and S. Trivedi,
``The Coulomb Branch of Gauge Theory from Rotating Branes,''
hep-th/9811120.}

\lr\Giddings{S. Giddings, ``The Boundary S-matrix and the AdS to
CFT Dictionary,'' hep-th/9903048.}

\lr\ceresole{A. Ceresole, G. Dall'Agata, R. D'Auria, and S. Ferrara,
``Spectrum of Type IIB Supergravity on $AdS_5\times T^{1,1}$: Predictions
On ${\cal N}=1$ SCFT's,'' hep-th/9905226.}

\lr\ms{J. Maldacena and A. Strominger,
``$AdS_3$ Black Holes and a String Exclusion Principle,''
hep-th/9804085. }
\lr\gks{
A. Giveon, D. Kutasov and N. Seiberg, ``Comments on String Theory
on $AdS_3$,'' hep-th/9806194. }


\baselineskip8pt
\Title{\vbox
{\baselineskip 6pt
{\hbox {PUPT-1863}}{\hbox{IASSNS-HEP-99/49 }}
{\hbox{hep-th/9905104}} 
{\hbox{   }}
}}
{\vbox{\vskip -30 true pt
\centerline {AdS/CFT Correspondence and Symmetry Breaking}
\medskip
\vskip4pt }}
\vskip -20 true pt 
\centerline{ Igor R. Klebanov}
\smallskip\smallskip
\centerline{Joseph Henry Laboratories, Princeton University, 
Princeton, New Jersey 08544}
\bigskip
\centerline  {Edward Witten}
\smallskip\smallskip
\centerline {Institute for Advanced Study, Olden Lane,
Princeton, New Jersey 08540 } 

\bigskip\bigskip
\centerline {\bf Abstract}
\baselineskip10pt
\noindent
\medskip
We study, using  the dual AdS description,
the vacua of field theories where some of the gauge symmetry is
broken by expectation values of scalar fields. 
In such vacua, operators built out of the
scalar fields acquire expectation values, and we show how to calculate
them from the behavior of perturbations to the AdS
background near the boundary. Specific examples include the ${\cal N}=4$
SYM theory, and theories on D3 branes placed on orbifolds and 
conifolds. We also clarify some subtleties of the AdS/CFT correspondence
that arise in this analysis.
In particular, we explain how scalar fields in AdS space of sufficiently
negative mass-squared can be associated with CFT operators of
{\it two} possible dimensions. All dimensions are bounded
from below by $(d-2)/2$; this is the unitarity bound for scalar
operators in $d$-dimensional field theory.
We further argue that the generating functional for correlators in the
theory with one choice of operator dimension is a Legendre transform
of the generating functional in the theory with the other choice. 

\bigskip
\Date {May 1999}

\def\Z{{\bf Z}}
\noblackbox \baselineskip 15pt plus 2pt minus 2pt 
\newsec{Introduction}

The AdS/CFT correspondence \refs{\jthroat,\GKP,\EW}
may be motivated by comparing stacks of
elementary branes with corresponding gravitational backgrounds in string
or M-theory. For example, the correspondence \kleb\ 
between a large number $N$ of coincident D3-branes and the 
3-brane classical solution leads, after
an appropriate low-energy limit is taken, to the duality between
${\cal N}=4$ supersymmetric $SU(N)$ gauge theory and Type IIB strings on
$AdS_5\times S^5$ \refs{\jthroat,\GKP,\EW}. 
This construction gives an explicit
realization to the ideas of gauge theory strings \refs{\GT,\AP}.

In order to construct the Type IIB duals
of other 4-dimensional CFT's, one may place the D3-branes at appropriate
conical singularities \refs{\DM,\Kehag,\KW,\FF,\MP}. 
Then the background dual to the CFT on the
D3-branes is $AdS_5\times X_5$ where $X_5$ is the Einstein manifold
which is the base of the cone.
Indeed, the metric of a 6-dimensional
cone $Y_6$ has the general form
\eqn\cone{ds_{\rm cone}^2 = d r^2 + r^2 ds_5^2\ . 
}
Here $Y_6$ is a cone over a five-manifold
$X_5$, and $ds_5^2$ is a metric on $X_5$.
If a large number $N$ of D3-branes is placed at the apex of the cone,
that is at $r=0$, then the resulting
geometry has the metric
\eqn\genmet{ ds^2 = H^{-1/2} (-dt^2 + dx_1^2 + dx_2^2 + dx_3^2) +
H^{1/2} ds_{\rm cone}^2 \ ,
}
where 
$$H = 1 + {L^4\over r^4} \ ,
\qquad\qquad L^4 \sim g_{\rm st} N (\alpha')^2 \ .
$$
In the near-horizon limit the
constant term in $H$ may be ignored and the geometry becomes 
$AdS_5\times X_5$
where $X_5$ is the base of the cone. Type IIB string theory in this
background is then conjectured to be dual to the infrared limit 
of the field theory on the stack of D3-branes. Some explicit examples
of such duality were exhibited in \refs{\KS,\LNV,\KW,\MP}.

In this paper we study in some detail 
the vacuum
states of these CFT's in which
some of the gauge symmetry is broken by expectation values of scalar 
fields.  In terms of the AdS description, such vacua arise
either by moving the threebranes away from the conical singularity
or from each other, or from the dynamics of the manifold $Y_6$,
whose singularity might be either resolved or deformed.
These cases are all somewhat similar as, whether the threebranes
are moved
or $Y_6$ is resolved or deformed, the threebranes tend
to end up at a smooth point in $Y_6$.
In fact, vacua obtained  by resolution, deformation, and threebrane
motion can all be described in an AdS language by using metrics
that look like $AdS_5\times X_5$ only near infinity.

These vacua  also fit in a common framework in the description
via boundary conformal field theory.   They are all obtained by
symmetry breaking, that is by giving expectation values to various scalar
fields.

Perhaps the simplest example of such gauge symmetry breaking arises
in the ${\cal N}=4$ SYM theory, by turning on scalar fields
such that the gauge group $SU(N)$ is broken down to
$S(U(N_1)\times U(N_2) \ldots \times U(N_k))$. In the language of
D3-branes, this corresponds to separating them into $k$ parallel stacks.
The appropriate geometry is the $k$-center threebrane solution
\refs{\jthroat,\klt}, and one may
once again take a scaling limit which amounts to dropping the constant
term in the Green's function $H$. Following \klt,
we will put the interpretation of the $k$-center solution
via gauge symmetry breaking on a more precise and systematic
basis  by using the general principles of the AdS/CFT correspondence
to compute expectation values of gauge-invariant order parameters in vacua
described by the $k$-center solution.

\def\R{{\bf R}}
A wider range of examples comes from considering threebranes near
a conical singularity.  The most elementary examples are
 the ``orbifold'' CFT's,  where
the six-dimensional cone is $\R^6/\Gamma$, with $\Gamma$  a discrete
subgroup of $SO(6)$ \refs{\DM,\KS,\LNV}. We denote the elements of $\Gamma$
as $\omega_i$.   If the D3-branes are displaced away
from the orbifold singularity to a transverse position $\vec y_0$, 
then the metric is given by
\eqn\orbmet{ ds^2 = H^{-1/2} dx^2 +
H^{1/2} (d\vec y)^2 \ ,
}
where the Green's function is 
$$H = L^4 \sum_{i=1}^n {1\over |\vec y - \omega_i \vec y_0|^4}\ ,
$$
and the resulting space is subsequently divided by $\Gamma$.
(Here we are denoting the four coordinates that parametrize the brane
world-volumes as $x$ and the six normal coordinates as $y$.)
Intuitively, if  $y_0$ is displaced from all of the orbifold
fixed points, then this metric has the same singular
structure as that obtained from $N$ D3-branes at a smooth point on
$\R^6$.  This suggests that the metric describes the flow
(via a Higgs effect) from an orbifold field theory \refs{\KS,\LNV} at short
distances to an $SU(N)$ theory with ${\cal N}=4$ supersymmetry
at long distances.  (For instance, if $\Gamma=\Z_n$, the orbifold
theory has gauge group $S(U(N)^n)$, which can be broken to a diagonal
$SU(N)$ by scalar expectation values.)
We will aim to put this interpretation on a precise and systematic basis 
by computing
the expectation values of the natural gauge-invariant order parameters.

Less elementary is the case that the conical manifold $Y_6$ is
not simply an orbifold.
A simple case is that $Y_6$ may be the conifold singularity in
complex dimension three, described in terms of complex variables
$w_1,\dots, w_4$ by the equation
$$ \sum_{a=1}^4 w_a^2 = 0
\ .
$$
The conifold admits a conical Calabi-Yau metric of the form
$$ ds^2 = d r^2 + r^2 d\Omega^2\ .  $$
Here 
\eqn\co{ d\Omega^2=
{1\over 9} \left(d\psi + \cos \theta_1 d\phi_1+ \cos \theta_2 d\phi_2\right)^2+
{1\over 6} \sum_{a=1}^2 \left(
d\theta_a^2 + \sin^2\theta_a d\phi_a^2 \right)
}
is the metric on the base of the cone \cd, which is
$T^{1,1}= (SU(2)\times SU(2))/U(1)$.
The ${\cal N}=1$
superconformal field theory with gauge group $SU(N)\times SU(N)$
that results when 
the D3-branes are placed at $r=0$, and is dual to $AdS_5\times T^{1,1}$,
was discussed in detail in \refs{\KW,\MP}. 

In this example, as anticipated above,
symmetry breaking can take several forms.
One may move the threebranes away from $r=0$ to a smooth point,
or one may resolve the singularity of $Y_6$ to get a smooth manifold
$Y_6''$ that looks like $Y_6$ near infinity (in which case the threebranes
are necessarily at a smooth point).  In either case, assuming
the threebranes are all at the same point, 
the low energy theory will be the ${\cal N}=4$ $SU(N)$ gauge theory.
Thus, the model analyzed in \refs{\KW,\MP} can flow in the infrared
to one of these vacua.
We will analyze the geometries that are relevant to these flows,
and compute the expectation
values of chiral superfields in these vacua, 
getting results that are in agreement with
field theory analysis \KW.

Section 2  of this paper is devoted
to some details of the AdS/CFT correspondence that will arise
in our analysis.  
In particular, we explain how scalar fields in AdS space of sufficiently
negative mass-squared can be associated with CFT operators of
{\it two} possible dimensions.
This subtlety is important for the conifold model,
because it contains such fields in its spectrum.
We also formulate, following similar ideas in
\refs{\BKL,\Vij,\klt,\BK,\Myers}, the general procedure 
for computing the expectation
value of an operator in a quantum vacuum that is related to a given
classical solution.  This will be important for our applications
to symmetry breaking. 
Those applications  are presented in section 3
for the ${\cal N}=4$ theory and orbifolds, and in section 4 for
the conifold.  Some technical details of the spectrum for the conifold
are given in an appendix.

\newsec{The Mass Spectrum And Operator Dimensions}

\subsec{Two Theories From The Same Lagrangian}

The AdS/CFT correspondence gives the following relation between 
the mass $m$ of a scalar in $AdS_{d+1}$ and the dimension $\Delta$ of the
corresponding operator \refs{\GKP,\EW},
\eqn\mop{\Delta (\Delta - d) = m^2\ .
}
There are two solutions, 
\eqn\branches{ \Delta_\pm = {d\over 2} \pm \sqrt{ {d^2\over 4} + m^2 }
\ ,
}
and it is often assumed that only $\Delta_+$ is admissible.
If true, this would imply that dimensions of scalar operators
are bounded from below by $d/2$, which is more stringent 
than the unitarity bound $(d -2)/2$.

In various explicit examples of the AdS/CFT duality, however,
 the field theory side contains operators of dimension less than
$d/2$. Examples of this include the large $N$ $(2,0)$
theory dual to M theory on $AdS_4\times S^7$ where one finds
operators of dimension $1$ \Aha, the F-theory constructions 
of $AdS_5$ duals \afm\ where one finds
dimensions $6/5$, $4/3$, $3/2$, and the D3-branes on the
conifold \KW\ where there are operators of dimension $3/2$. There
are also operators in the D1-D5 system 
with arbitrarily low dimensions \refs{\ms,\gks}.
In all these examples the supersymmetry unambiguously requires the
presence of these low dimensions,
all of which are consistent with the unitarity bound but
are smaller than $d/2$.
Therefore, if the AdS/CFT correspondence is correct, then there must
be a loophole in the conclusion that only $\Delta_+$ is admissible.
This issue was raised and discussed in \BKL, where the relevance
of old work by Breitenlohner and Freedman \BF\ was also suggested.

Breitenlohner and Freedman considered a free scalar field of mass $m$
in AdS space, and showed that, while for $m^2>-{d^2\over 4}+1$ there is
a unique admissible boundary condition for such a field that is
invariant under the symmetries of AdS space, leading to
a unique AdS-invariant quantization, for
\eqn\range{ -{d^2\over 4} < m^2 < - {d^2\over 4} + 1
}
there are two possible quantizations.
These two possibilities correspond to the fall-off
of scalar wave functions as $z^{\Delta_+}$ and $z^{\Delta_-}$ near
$z=0$, where the $AdS_{d+1}$ metric is written as
\eqn\adsmet{ ds^2 ={1\over z^2} \left (dz^2 + \sum_{i=1}^d (d x^i)^2
\right )
\ ,
}
and we set $L=1$.
Breitenlohner and Freedman formulated their arguments in Hamiltonian
terms and looked for boundary conditions that make the energy finite.
Instead of repeating their argument, we will give a heuristic derivation
of the result in Euclidean space, by requiring finiteness of the action.

The conventional expression for the Euclidean action of a massive scalar field
in $AdS_{d+1}$ is
\eqn\action{ {1\over 2} \int d^{d+1} x \sqrt g
\left [ g^{\mu\nu} \partial_\mu \phi \partial_\nu \phi
+ m^2 \phi^2 \right ]=
{1\over 2} \int d^d x dz  z^{-d+1}\left [(\partial_z \phi)^2+
(\partial_i\phi)^2  + {m^2\over z^2} \phi^2 \right ]
\ .
}
Solutions of the classical equations of motion of this theory behave
near $z=0$ -- that is, near the boundary of AdS space -- as
\eqn\bound{\phi (z, \vec x) \rightarrow z^\Delta (A(\vec x) + O(z^2)),}
where $\Delta$ can be either $\Delta_+$ or $\Delta_-$.  Any boundary
condition on the field must set to zero half of the modes of the field
near the boundary.  It is natural and completely AdS-invariant
to pick a particular root, $\Delta=\Delta_+$
or $\Delta=\Delta_-$, and require that $\phi$ behave as in \bound\ near
the boundary.  (Of course, we do not require that $\phi$ obey
the classical equations of motion in the interior of AdS space.)
With this asymptotic condition, 
 the action \action\ is finite for $\Delta> d/2$.
But the bound on $\Delta$ 
can be relaxed by adding appropriate boundary terms
to the action. 
By integrating by parts and discarding the boundary term, 
we can replace the action by
\eqn\newaction{
{1\over 2} \int d^{d+1} x \sqrt g
\ \phi (-\nabla^2 + m^2) \phi \ . }
The boundary term in this integration by parts is nonzero (and in fact
divergent) if $\Delta \leq d/2$, so in writing the action \newaction,
we are modifying the definition of the action.
The modified action integral is convergent if 
\eqn\jugglo{\Delta>{d-2\over 2}.}
This is  precisely the unitarity bound on the dimension of a scalar
operator in $d$ dimensions, so in particular we cannot expect by any
further device to get even smaller $\Delta$'s.
 In the mass range
\range, this condition allows  $\Delta=\Delta_-$ as well as $\Delta_+$,
while for larger $m^2$ only $\Delta=\Delta_+$ is allowed.

Though convergent for fields that obey the boundary
conditions, the action \newaction\ is not
manifestly positive definite.  However, it is positive definite,
because the operator $-\nabla^2+m^2$ is positive definite
for the range of $m^2$ of interest, namely $m^2>-d^2/4$.

Thus, as pointed out in \BF, there are two different  AdS-invariant
quantizations of the scalar field with $m^2$ in the range \range.
One Lagrangian -- in this case that of a scalar field of given $m^2$ --
can give rise to two different quantum field theories in AdS space,
depending on the choice of boundary condition.
According to the general AdS/CFT correspondence,  any quantum
field theory in AdS space is equivalent to a conformal field theory
on the boundary.  The two different AdS theories with a given $m^2$
will correspond to two different CFT's, one with an operator of dimension
$\Delta_+$ and the other with an operator of dimension $\Delta_-$.
\foot{Of course, treating the AdS scalar field as a free field of mass $m^2$
can never be precisely right, since this field always interacts at least
with gravity.  There will in general
be mass renormalization, in which case the dimensions
of the operators will not be precisely $\Delta_+$ and $\Delta_-$.}
In many examples, one of the two theories is much more readily studied
than the other because one is supersymmetric and the other is not.
But both exist in principle.

\subsec{Correlation Functions}

Our main remaining goal will be to define the correlation functions
from the AdS/CFT correspondence for both choices of the theory.
In the process we will also give an important formula for the expectation
values of operators.
We will look for the definition of the Euclidean 
action which generates properly normalized correlation functions. 

To compute correlation functions, one must relax the boundary condition
\bound, so we have to exercise additional care in defining the action.
Indeed, there is a subtlety with the normalization of the two-point
function in the AdS/CFT correspondence.
In \FMMR, it was shown that an extra factor of $(2\Delta - d)/d$, not 
coming in an obvious way from evaluating the classical action,
is needed for consistency with the Ward identities.
This factor was then derived by imposing the boundary condition at
$z=\epsilon$, as advocated in 
\refs{\GKP,\MW,\FMMR}, and taking the $\epsilon\rightarrow 0$
limit at the end of the calculation.
We will present a different way of obtaining this factor which
involves adding an appropriate boundary term to the action.

In calculating correlation functions of vertex
operators from the AdS/CFT correspondence, 
the first problem is to reconstruct an on-shell field in $AdS_{d+1}$
from its boundary behavior. If $\Delta$ is one of the roots of \mop,
then one requires that for small $z$
\eqn\bc{\phi (z, \vec x)\rightarrow z^{d-\Delta} 
[\phi_0 (\vec x) + O(z^2)]
+z^\Delta [A(\vec x) + O(z^2)] \ ,
}
where $\phi_0 (\vec x)$ is a prescribed ``source'' function and 
$A(\vec x)$
describes a physical fluctuation that will be determined from the source
by solving the classical equations.
In our discussion so far, we only considered the physical 
fluctuation $A(\vec x)$.

We begin with the usual case $\Delta = \Delta_+$.  In this case,
the first term in \bc\ dominates over the second near $z=0$, and
the construction of $\phi (z, \vec x)$ from $\phi_0(\vec x)$
is usually accomplished with the help of the bulk-to-boundary
propagator \refs{\EW,\FMMR},
\eqn\propa{ K_\Delta (z, \vec x, \vec x') = \pi^{-d/2} {\Gamma (\Delta)\over
\Gamma (\Delta- (d/2))}
{z^\Delta\over (z^2 + (\vec x-\vec x')^2)^\Delta }
\ ,
}
so that
\eqn\bulk{ \phi(z, \vec x) =
\int d^d x' K_\Delta(z, \vec x, \vec x') \phi_0 (\vec x')
\ .
}
The normalization in \propa\ is chosen so that \bc\ is satisfied.
We note that 
$$A(\vec x)= \pi^{-d/2} {\Gamma (\Delta)\over \Gamma (\Delta- (d/2))}
\int d^d x' \phi_0 (\vec x') |\vec x-\vec x'|^{-2\Delta}
\ .$$
For extended sources this is a formal expression because it diverges 
for $\Delta > d/2$, but
it will be useful after appropriate regularization is taken into account.
We may also consider a localized source, $\phi_0 (\vec x) = \delta^d
(\vec x-\vec x')$. Then
\eqn\localsource{A(\vec x)= \pi^{-d/2} {\Gamma (\Delta)\over \Gamma (\Delta- 
(d/2))}
|\vec x-\vec x'|^{-2\Delta}
}
and it was observed in \Vij\ that, up to a
normalization factor, this is the two-point
function $\vev{\O(\vec x) \O(\vec x')}$.
This suggests that $A(\vec x)$
has the
interpretation of the expectation value of the operator $\O(\vec x)$
in the theory where another operator $\O$ is inserted at $\vec x'$.
We will see that the precise relation is
\eqn\expect{A(\vec x) = {1\over 2\Delta -d} \vev{\O(\vec x)}\ 
}
Up to normalization this is the same relation as the one advocated
in \refs{\BKL,\Vij,\klt,\BK,\Myers}.
The precise factor is related to the normalization
of the two-point function first found in \FMMR.
We will be able to show that this relation holds beyond the linearized
approximation.

In order to define the value of action on the solution
\bulk, it is convenient to
introduce another field, $\chi$, through
$$ \phi( z, \vec x) = z^{d-\Delta} \chi(z, \vec x)
\ .
$$
After integrating by parts 
and discarding an appropriate boundary term, the
action assumes the following form:
\eqn\newact{ I={1\over 2} \int d^d x dz\ z^{d+1- 2\Delta}
\left [ (\partial_z \chi)^2 + (\partial_i \chi)^2
\right ]
\ .}
We propose to define the two-point function for $\Delta=\Delta_+$
using this action.
It differs from the original action \action\ in that the leading
small $z$ divergence has been discarded.  The action integral in
\newact\ is convergent if $d/2+1> \Delta >d/2$.  For $\Delta\geq d/2+1$,
a more complicated subtraction of boundary divergences is needed to
get a well-defined action.  This corresponds to the fact that the 
conformal field theory generating
functional that we will compute has additional short distance singularities
if $\Delta \geq d/2+1$.  We will not explicitly make the additional
regularization of the action 
that is needed for $\Delta$ in this range.

Now we are ready to calculate the two-point function from the AdS/CFT
correspondence. We need to evaluate the improved action $I$ in terms of
$\phi_0 (\vec x)$; that is, we need to evaluate $I$ for a classical
solution \bulk\ with 
given $\phi_0(\vec x)$. Integrating \newact\ by parts we find
\eqn\twopoint{ I= - \lim_{z\to 0} z^{d+1 - 2 \Delta}
\int d^d \vec x {1\over 2}\chi \partial_z \chi
\ .
}  
In evaluating this expression, 
the $\phi_0\cdot \phi_0$ terms vanish if $\Delta< d/2+1$,\foot{To prove this,
one uses the fact that the correction
to the $\phi_0$ term in a classical solution \bc\ is of order $z^2$.}
and the $A\cdot A$ terms vanish
if $\Delta>d/2$.  To be more precise, to evaluate \twopoint, we can replace
$\chi$ by $\phi_0$ and 
 $\partial_z \chi$ by
$(2\Delta-d) z^{2\Delta -d -1} A (\vec x)$ for small $z$.  (Terms with
$A$ coming from $\chi$ and $\phi_0$ from $\partial_z\chi$ vanish using
the property emphasized in the last footnote.) So  we find
that
\eqn\twofin{ I(\phi_0) =- (\Delta- (d/2)) \pi^{-d/2}
{\Gamma (\Delta)\over \Gamma (\Delta-
(d/2))}
\int d^d \vec x \int d^d \vec x'
{\phi_0 (\vec x) \phi_0 (\vec x')\over |\vec x- \vec x'|^{2\Delta} }
.
}
In particular, as expected, $\Delta$ is the  dimension of the operator
${\cal O}$ that couples to the source $\phi_0$ in the boundary conformal
field theory.  
Because of the divergence for $\vec x$ near $\vec x'$ \twofin\ has
to be understood in an appropriately regularized sense.
For example, the corresponding expression in momentum space is
\eqn\Four{I(\phi_0)= -{1\over 2} \int {d^4 k\over (2\pi)^4}
\phi_0 (k) \phi_0 (-k) f_+ (|k|)
\ ,
}
where
\eqn\twoFour{f_+ (|k|)= - 2\nu \left ( {|k|\over 2}\right )^{2\nu}
{\Gamma(1-\nu)\over
\Gamma (1+\nu)}
}
is the Fourier transform of the two-point function 
(we have defined $\nu = \sqrt {m^2 + d^2/4}=\Delta_+ -d/2$).
\Four\ is finite for $\phi_0(k)$ that fall off
sufficiently fast for large $k$.
Note that the Fourier transform of
$|\vec x|^{-2\Delta_+}$ is actually UV divergent and when appropriately
defined has the negative coefficient that is indicated.

We will not attempt a similar derivation in detail for
$\Delta>d/2+1$. However, we claim that in this range, after the additional
subtractions of boundary terms that are needed to make $I$ finite, 
the $\phi_0\cdot \phi_0$ terms vanish and the $\phi_0\cdot A$ terms can be
evaluated to give the same formula as \twofin.   

The overall minus sign in \twofin\ is crucial: 
since $\phi_0$ is interpreted as the source coupling
to the CFT operator ${\cal O}$, this is the correct sign to insure
the positivity of the two-point function.
Indeed, $\exp(-I)$ is interpreted in the boundary field theory
as $\langle \exp(\int \phi_0{\cal O})\rangle$.  
So negativity of \twofin\ is needed
for positivity of the correlation function 
$\langle {\cal O}(\vec x){\cal O}(\vec x')\rangle$.

\bigskip\noindent
{\it One-Point Function In Presence Of Sources}

The prefactor $\Delta- (d/2)$ in \twofin\ is also important:
this is the factor advocated in \FMMR. Due to the presence of this
factor, we see, on comparing \localsource\ to \twofin, that the relation
$\vev {\O(\vec x)} = (2\Delta - d) A(\vec x)$ holds to linear order
in the sources.
Let us show that this relation holds to all orders.

\bigskip
\centerline{\vbox{\hsize=4.0in
\centerline{\psfig{figure=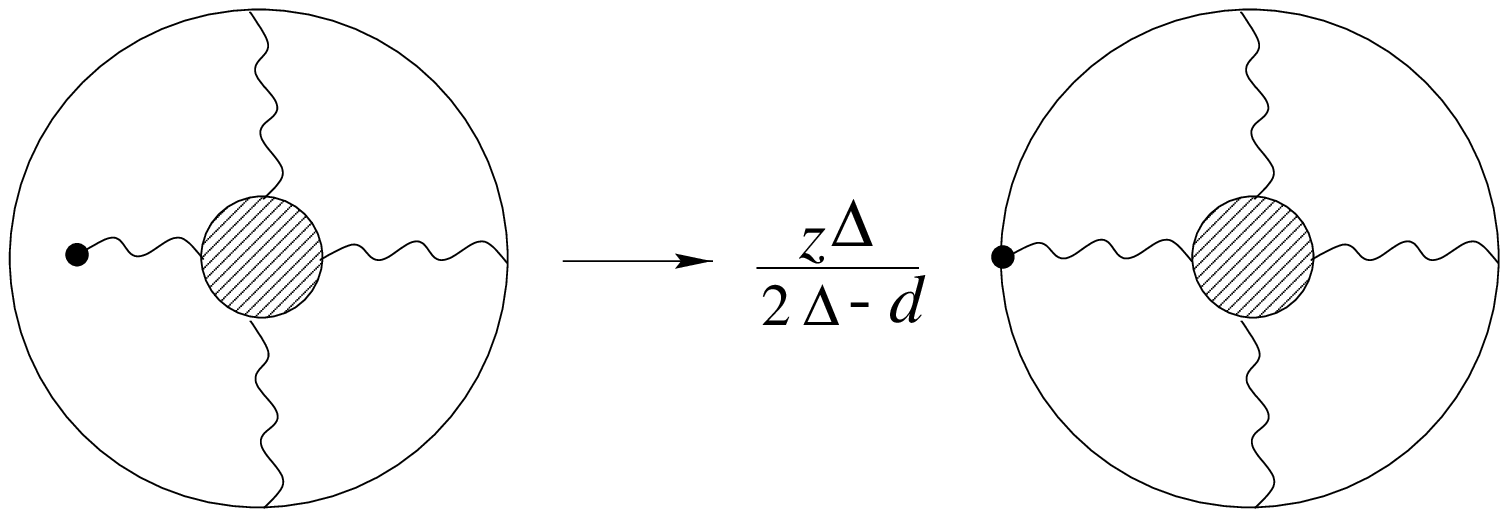}}
\vglue.4in
{\tenpoint
The diagrammatic relation between the term 
$A(\vec x)z^\Delta$ in the asymptotic
behavior of the field in an arbitrary correlation function or physical
process (left) and the same process with an extra insertion of an
operator $\O(\vec x)$ (right).
The solid black dot  represents on the left
a point in AdS near the boundary where a field is measured
 and on the right a  point on the boundary at which an
operator is inserted.  As the black dot approaches the 
boundary, the two figures are related by the indicated factor.}
}}
\bigskip

The expectation value of $\O(\vec x)$ is given by
the sum over diagrams where a bulk-to-boundary propagator
$K_\Delta (\vec x; z',\vec x')$
connects the point $\vec x$ to the rest of the diagram with source
points located at the boundary. The classical field $\phi(z,\vec x)$
is given by summing the same diagrams, except the bulk-to-boundary
propagator is replaced by the bulk-to-bulk propagator leading
to the point $(z,\vec x)$. The normalized expression for 
the bulk-to-bulk propagator is given, for instance, in \DFR:
$$ G_\Delta(z, \vec x; z', \vec x')=
{\Gamma(\Delta) \Gamma (\Delta- {d\over 2} +{1\over 2})
\over (4\pi)^{(d+1)/2}
\Gamma(2\Delta - d+1)} (2 u^{-1})^\Delta 
F(\Delta, \Delta - {d\over 2} +{1\over 2}; 2\Delta - d+1; -2 u^{-1})\ ,
$$
where $F$ is the hypergeometric function, and
$u= [(z-z')^2 + (\vec x-\vec x')^2]/2 z z'$.
We note that, as $z\rightarrow 0$ away from the source
points, $\phi(z,\vec x)\rightarrow z^\Delta
A(\vec x)$. In this limit
$$ G_\Delta (z, \vec x; z', \vec x')
\rightarrow z^\Delta {K_\Delta (\vec x; z',\vec x')\over 2\Delta-d}
\ .
$$
This property of the normalized Green's functions, first emphasized in
\Giddings, thus provides
a general explanation of the relation \expect.

\bigskip\noindent
{\it Extension To $\Delta<d/2$}

The derivations presented so far apply to the theory with $\Delta=\Delta_+$.
As we have explained, for masses in the range \range\ it should be
possible to define a different theory where the operator 
corresponding to the scalar field in $AdS_5$ has dimension
$\Delta_-$. The $\Delta_-$ theory is not independent from the $\Delta_+$
theory but is, in fact, related to it by a canonical transformation that
interchanges the roles of $\phi_0(\vec x)$ and $A(\vec x)$.
The fact that $\phi_0$ and $A$ are conjugate variables is also suggested
by the group-theoretic analysis in \Dobrev. If from the point
of view of the $\Delta_+$ theory $\phi_0$ is ``the source'' and 
$(2\Delta-d) A$
is ``the field,'' then the opposite is true for the $\Delta_-$ theory.
This strongly suggests that the generator of connected
correlators of the $\Delta_-$ theory is obtained by
Legendre transforming the generator of connected
correlators of the $\Delta_+$ theory.
This type of relationship is familiar from the Liouville theory
where it has been suggested that 
theories with two different branches of gravitational dressing
of a given operator are related by a Legendre transform \KH.

To see how this works for the two-point functions, it is convenient to
use Fourier space. The quadratic part of the
action is given in \Four, where $f_+(|k|)$ is the 
Fourier transform of the two-point function in
the $\Delta_+$ theory.
The Legendre transform is carried out by first setting
$$ J(\phi_0,A)=I(\phi_0) +(2\Delta-d)\int 
{d^4k\over (2\pi)^4}\phi_0(k)A(-k).$$
We have included a factor of $(2\Delta -d)$ based on the idea that the
conjugate of $\phi_0$ is actually $(2\Delta-d)A$.  As we will see, this
factor, though we have not justified it precisely, 
gives the nicest normalization
for the two point function of the transformed theory. 
The Legendre transformed
functional $\tilde I(A)$ is the minimum of $J(\phi_0,A)$ 
with respect to $\phi_0$
(for fixed $A$), and is explicitly
$$\tilde I(A)= -{1\over 2} \int {d^4 k\over (2\pi)^4}
A(k) A (-k) f_- (|k|)
$$
with
\eqn\Lrel{ f_- (|k|)= -{(2\Delta - d)^2 \over f_+ (|k|)}
\ .
}
Substituting \twoFour\ we find
$$ f_- (|k|)= 2\nu \left ( {|k|\over 2}\right )^{-2\nu} {\Gamma(1+\nu)\over
\Gamma (1-\nu)}
\ ,$$
which is related to $f_+ (|k|)$ by $\nu\rightarrow - \nu$.
Fourier transforming back to position space (via an integral which now
converges, so that there is no additional flip of sign) and recalling
that the two-point function is minus the second derivative of the effective
action, we find that in the $\Delta_-$ theory 
\eqn\newtwo{
\vev{{\cal O}(\vec x){\cal  O}(\vec x')}=
{(2\Delta_- - d) \Gamma (\Delta_-)\over \pi^{d/2} \Gamma (\Delta_- -
(d/2))} {1\over 
|\vec x- \vec x'|^{2\Delta_-} }
\ .
}
Happily, this function is indeed positive for all dimensions $\Delta_-$
above the unitarity bound. Note also that \newtwo\ is exactly the same
formula as the one we would find by extrapolating \twofin\ to
$(d-2)/2 < \Delta < d/2$.

One may be puzzled by the double zero of the two-point function
at $\Delta=d/2$. In fact, value of $\Delta$, which corresponds to
$m^2=-d^2/4$ is the special case where $\Delta_+ = \Delta_-$.
Here the two possible small $z$ behaviors of a classical
solution are $z^{d/2}\ln (z/z_0)$ and
$z^{d/2}$. 
Now, only one conformally-invariant boundary condition is
possible.  We can ask that physical fluctuations behave as $z^{d/2}$
with no $z^{d/2}\ln(z/z_0)$ term.  But it does not make sense
to ask that they behave as $z^{d/2}\ln (z/z_0)$ with no $z^{d/2}$ term;
such a condition would depend on the choice of $z_0$, violating conformal
invariance.  There is therefore likewise only one natural way
to incorporate an external  source $\phi_0$; we require that 
$\phi(z,\vec x)$ approaches $\phi_0 (\vec x) z^{d/2}\ln (z/z_0)$
for small $z$.\foot{Once such a source is included, the $\ln (z/z_0)$
term gives a violation of conformal invariance in defining the
expectation value $\langle{\cal O}\rangle$.  This is the correct
answer from the point of view of conformal field theory.  If ${\cal O}$
has dimension $d/2$, there is a logarithmic divergence in the two point
function $\int d^dx e^{i\vec p\cdot \vec x}
\langle{\cal O}(\vec x){\cal O}(0)\rangle$, as a 
result of which if one computes $\langle{\cal O}\rangle$ in the presence
of a source, there is a logarithmic  violation of conformal invariance.}
The bulk field as defined by \bulk, however, behaves 
for $\Delta=d/2$ as $z^{d/2}\phi_0 (\vec x)$. To remove this discrepancy
we may simply divide the operator ${\cal O}$ 
 by $\Delta- (d/2)$, in which case we should  divide and multiply $\phi_0$
and $A$ by the same factor.  With this perhaps more useful normalization,
the AdS two-point function becomes
\eqn\twomod{ \vev {\O(\vec x) \O(\vec x')} = 2 \pi^{-d/2} 
{\Gamma (\Delta)\over \Gamma (\Delta-
(d/2)+1)}
{1\over |\vec x- \vec x'|^{2\Delta} }
\ ,
}
which does not vanish until the dimension approaches the unitarity
bound. This field renormalization has a similar effect on the three-point
functions: the leg factors ${1\over \Gamma (\Delta- (d/2))}$
that appear in the results of \FMMR\ are changed into
${1\over \Gamma (\Delta- (d/2)+1)}$ so that the zeroes at
$\Delta=d/2$ are eliminated. This is in accord
with the field theory results
which give non-vanishing correlators of dimension $d/2$ operators,
such as those found in the ${\cal N}=4$ SYM theory \Lee.
In general, one should keep in mind that the AdS results agree
with field theory calculations only after certain dimension dependent
field normalization factors are included \Lee.

To summarize the present section, in the range of scalar masses
\range, there are two possible theories. One of them associates to the scalar
field
an operator of dimension $\Delta_+$, while the other associates
an operator of dimension
$\Delta_-$. The two different definitions of the theory correspond
to interchanging the source $\phi_0(\vec x)$ and the expectation value
$(2\Delta-d) A(\vec x)$ defined by the boundary behavior \bc. 
Thus, the generating functionals of correlation functions are
related by a Legendre transform. This is analogous to the situation
found in Liouville theory where the generating functional corresponding
to the theory with one branch of gravitational 
dressing is the Legendre transform
of the generating functional corresponding to the other branch \KH.

\newsec{Examples of symmetry breaking}

In this section we discuss perhaps the simplest example
of the AdS/CFT duality in presence of symmetry
breaking. This example was discussed previously in \refs{\jthroat,\klt}:
the gauge symmetry present on coincident
D3-branes in flat space may be broken simply by separating them
into several parallel stacks. Below we will discuss a simple case of breaking
$SU(N)$ down to $S(U(N_1)\times U(N_2))$ by separating $N$ coincident
D3-branes into two parallel stacks containing $N_1$ and
$N_2$ coincident D3-branes. Its generalizations to
more complicated breaking patterns will then be immediate.

The two-stack
 configuration of branes corresponds to Higgsing of the ${\cal N}=4$
gauge theory by scalar fields
\eqn\utury{\vec X=\left(\matrix {\vec d_1\cdot I_1 & 0 \cr 0 & \vec d_2\cdot
I_2 \cr}\right)
}
where $\vec d_i$ is the position of the $i$-th brane stack
and $I_i$ is the $N_i\times N_i$ identity matrix.
Such a Higgsing gives expectation values to the chiral fields
${\cal O}^{(n)}_{i_1i_2\dots i_n} = \Tr\,X_{i_1}X_{i_2}\dots X_{i_n}-{\rm
trace~terms}$:
\eqn\forgetful{\langle {\cal O}^{(n)}_{i_1i_2\dots i_n}\rangle\sim 
N_1 [(d_1)_{i_1}(d_1)_{i_2}\dots (d_1)_{i_n} - {\rm traces}]
+ N_2 [(d_2)_{i_1}(d_2)_{i_2}\dots (d_2)_{i_n} -{\rm traces}]\ .} 
We wish to see how these order parameters emerge in the AdS description.

To obtain that description, we must find the appropriate analog
of the $AdS_5\times S^5$ metric. For this, we proceed along lines
described in the introduction.  
The Green's function with two separated sources is
\eqn\Gr{ H=L^4 \left({a_1\over |\vec y-\vec d_1|^4} + 
{a_2\over |\vec y - \vec d_2|^4} 
\right)
\ ,
}
where $y_1,\dots,y_6$ are the coordinates normal to the brane and
\eqn\charges{ a_i={N_i\over N}
\ .
}
It is always possible, by a shift
of the coordinates, to choose the origin at the ``center of mass''
so that $a_1 \vec d_1+ a_2\vec d_2=0$. Adopting this choice we find
that the Green's function \Gr\
can be given the following Taylor series expansion at large $\vec y$
($r=|\vec y|$): 
\eqn\ruty{H={L^4\over r^4} 
\left(1+\sum_{n=2}^\infty 2^{n} {(n+1)
d^{(n)}_{i_1i_2\dots i_{n}}y^{i_1}y^{i_2}\dots y^{i_{n}}\over
r^{2n}} \right ),
}
with 
\eqn\rep{
d^{(n)}_{i_1i_2\dots i_{ n}}=
a_1 [(d_1)_{i_1}(d_1)_{i_2}\dots (d_1)_{i_n} - {\rm traces}]
+ a_2 [(d_2)_{i_1}(d_2)_{i_2}\dots (d_2)_{i_n} -{\rm traces}]\ .} 
The trace terms are precisely such that $d^{(n)}_{i_1i_2\dots i_{ n}}$
is a traceless symmetric tensor.

Given the Green's function, the corresponding spacetime metric is,
as in \genmet,
\eqn\furry{ds^2=H^{-1/2}(-dt^2+dx_1^2+dx_2^2+dx_3^2)+H^{1/2}\sum_{j=1}^6 
dy_j^2.}
The large $r$ behavior can be worked out using \ruty.  In particular,
if we simply set $H=L^4/r^4$, we get the familiar $AdS_5\times S^5$
metric, with as usual $r$ combining with $t,x_1,x_2,x_3$ to make
the five $AdS_5$ coordinates.  In this description, the boundary of
$AdS_5$ is at $r=\infty$, and $r$ is related to the parameter $z$ of 
section 2 by
\eqn\yuggo{r={L^2\over z}.}
Expanding $H^{1/2}$ to linear order in
$a_i$ we find that for every $n=2,3,\dots$, there
is in the metric a correction term proportional to
\eqn\ikko{ {d^{(n)}_{i_1\dots i_{n}}\hat y^{i_1}\dots \hat y^{i_n}
\over r^{n}} \ . }
where, in general,
$d^{(n)}_{i_1\dots i_n}$ are the tensors that appear in the partial wave
expansion of the Green's function $H$.

As we have explained in eqn. \expect, the existence of a correction
to the metric proportional to $r^{-n}=z^{n}$ means that
a conformal field of dimension $n$ has an expectation value.
Given the structure of \ikko, this conformal field clearly transforms
under the $SO(6)$ symmetry of the $y^i$ as 
$d^{(n)}_{i_1\dots i_{n}}$.  The field in question
is precisely ${\cal O}^{(n)}_{i_1i_2\dots i_{n}}$ and, up to
proportionality constants that we have not checked, 
the $d^{(n)}_{i_1\dots i_{n}}$ given in \rep\ agree with the
expectation values \forgetful\ calculated in field theory.
In fact, in the AdS/CFT correspondence,
 ${\cal O}^{(n)}_{i_1i_2\dots i_n}$ is mapped to conformal
fluctuations in the metrics of $AdS_5$ and $S^5$ (and a related fluctuation
in the four-form gauge potential); in the present context,
such a fluctuation comes  from the $r^{-n}$ correction
term in $H$.

In a more general multi-stack example, with
\eqn\huffov{
H=L^4\sum_i {a_i\over |\vec y-\vec d_i|^4},\qquad \sum_i a_i=1,}
we can always eliminate the spin one or ``dipole'' harmonic in the 
expansion of $H$ by adding a constant to $\vec y$ to transform
to a ``center of mass'' frame with $\sum_i a_i\vec d_i=0$.  The
higher harmonics cannot be so eliminated and obey no general restrictions,
so in general we get expectation values of the chiral fields
${\cal O}^{(n)}$ for all $n\geq 2$. 
The vanishing of ${\cal O}^{(1)}$ means that the classical
supergravity solutions of this kind, modulo coordinate
transformations, are in natural
correspondence with the vacua of an $SU(N)$, rather than $U(N)$,
gauge theory.

So far we have discussed the terms in the metric coming from
expanding $H^{1/2}$ to linear order in the ``charges'' $a_i$.
However, as pointed out in \klt, starting with $n=4$ the coefficients
of $r^{-n}$ terms in the metric also have corrections containing
higher powers of $a_i$. Since $a_i=N_i/N$ the structure of these
terms is suggestive of expectation values of operators  containing
more than a single trace \klt. For example, for $n=4$, in addition to
$\vev{{\cal O}^{(4)}}$ we also seem to find ${1\over N} \vev{{\cal O}
^{(2)} {\cal O}^{(2)}}$.
We postpone investigation of these extra terms for the future.

\bigskip\noindent{\it Orbifolds}

Orbifolds obtained by dividing the space $\R^6$ 
transverse to the threebranes
by a finite subgroup $\Gamma$ of $SO(6)$
can be discussed in a very similar fashion.  
If all branes are at the origin in $\R^6$, we get a theory whose
infrared limit is described by $AdS_5\times S^5/\Gamma$.  For
example, in some much-studied cases with $\Gamma={\bf Z}_n$ 
\refs{\KS,\LNV},
the $AdS_5\times S^5/\Gamma$ geometry is dual to a gauge
theory with gauge group $S(U(N)^n)$.

Higgsing of these theories will be described in the AdS/CFT correspondence
by replacing $AdS_5\times S^5/\Gamma$ by a solution that looks like
that near the boundary but is different in the interior.
For instance, Higgsing of the gauge theory to a diagonal $SU(N)$ is
described in terms of branes by placing all branes at the same smooth
point in $\R^6/\Gamma$, away from all orbifold singularities.
The corresponding Green's function is
\eqn\jugi{H=L^4\sum_{i=1}^n{1\over |\vec y-\omega_i\vec y_0|^4},}
with generic $\vec y_0$ and with 
$\omega_i$ the elements of $\Gamma$.  Again, $H$ can be
expanded in spherical harmonics. The leading term at long distances
is $H=nL^4/r^4$, and gives back an $AdS_5\times S^5/\Gamma$ metric near the
boundary; the corrections vanish as higher powers of $r$ and correspond
to expectation values of chiral fields in the gauge theory.
Other symmetry-breaking vacua are described, in the AdS language,
by taking $H$ to have sources at fixed points of some of the elements
of $\Gamma$ or to have sources consisting of
more than a single $\Gamma$ orbit.

Note that if the group $\Gamma$ acts on $\R^6$ with no invariant vectors,
then a ``dipole'' term, of order $1/r^5$, is always absent in the expansion
of $H$ in spherical harmonics,
as this term is annihilated by the sum over $\omega_i$. 
Even if there are invariants in the action of $\Gamma$ on $\R^6$, the
dipole term can be eliminated by shifting to ``center of mass'' coordinates
in the directions on which $\Gamma$ acts trivially.  Once this is done,
 there is
no $1/r^5$ term in $H$, so we never get an expectation value for a chiral
field of dimension 1.  This is just as well, for in a unitary quantum
field theory in four dimensions, a scalar field of dimension 1 must be
free and so cannot be described by the dynamics in the bulk of AdS space.
Such a field can only enter the AdS/CFT correspondence as a ``singleton''
field supported at infinity.
All assertions in this paragraph remain true if the configuration of
sources in \jugi\ is generalized to a more general $\Gamma$-invariant
configuration.

\bigskip\noindent{\it Global Structure}

To conclude this section, we will make a few remarks about the global
structure of the supergravity solutions that we have discussed.

We recall the form of the $AdS_5 \times S^5$ solution:
\eqn\ugurfy{  ds^2 = {L^2\over r^2}dr^2+
{r^2\over L^2} \sum_{i=1}^4 dx_i^2 + L^2 d\Omega_5^2.}
Here $\Omega_5$ is the metric on a round five-sphere,
and $x_i, $ $i=1,\dots, 4$, are coordinates on the four dimensions
parallel to the threebranes.

Roughly speaking, the boundary of $AdS_5$ is at $r=\infty$
(or $z=0$ in the notation of section 2, the relation being
$z=L^2/r$).  The boundary thus appears to be a copy
of $\R^4$, parametrized by the $x_i$.  However, this is not the whole story, 
because the coordinate
system used in \ugurfy\  behaves badly at $r=0$.  In fact,
$r=0$ contributes one more point to the boundary of $AdS_5$
(heuristically, because the coefficient of $dx_i^2$ vanishes at
$r=0$, $r=0$ is just a single point on the boundary).  When we add
this point, the boundary of $AdS_5$ is compactified from $\R^4$ to $S^4$.

The compactness of $S^4$ means that, as long as we only consider
perturbations that admit this same global structure at infinity,
we cannot encounter infrared divergences.  Moreover, the symmetry breaking
or Higgsing phenomena that we have studied above cannot occur if
the boundary is $S^4$.  One way to explain this uses the positivity of the
scalar curvature $R$ of $S^4$.  The scalar fields $X$ of the ${\cal N}=4$
theory have a conformally invariant quadratic term in the action,
\eqn\polko{\int d^4x \sqrt g \left(\Tr \,(dX)^2+
{R\over 6}\Tr \,X^2\right)\ ,}
that is strictly positive definite and prevents $X$ from acquiring
an expectation value, as long
as one only considers perturbations that make sense when the boundary
is $S^4$.  (In particular, a constant $X$ field on $\R^4$ has a singularity
at the ``point at infinity'' if one tries to compactify to $S^4$.)

Because of all these facts, the theory on $S^4$ is completely stable
under sufficiently small perturbations, and there is in fact a well-defined
procedure for computing its correlation functions.  
If we want to study symmetry-breaking, we cannot achieve this
degree of ``safety''; we have to allow perturbations that cannot
be naturally interpreted on $S^4$ and for which the metric
will, in fact, have dangerous and difficult to interpret regions.
To see how this works, we rewrite \ugurfy\ in the form \furry\
with 
\eqn\bugufy{H={L^4\over r^4}.}
This metric appears to be badly behaved near $r=0$, but as we have
discussed, this is part of the one-point compactification that converts
the boundary from $\R^4$ to $S^4$.  Now, however, consider
one of the symmetry-breaking choices like 
\eqn\bugurfy{H=L^4\left({a_1\over |\vec y-\vec d_1|^4}+
{a_2\over |\vec y-\vec d_2|^4}\right).}
There are now two dangerous points, at $\vec y=\vec d_1,\ \vec d_2$.  Either
of these looks in this coordinate system just like the $r=0$ point
in the case \bugufy. With two singularities, it is not possible to absorb
them as part of the ``boundary,'' since there is no reasonable
compactification of $\R^4$ by adding two points.  Thus, for the 
symmetry-breaking
vacua, we are committed to thinking of the boundary as $\R^4$. 
This is just another manifestation of the fact that positivity of
\polko\ makes symmetry breaking impossible if the boundary is $S^4$. 

With the boundary, on which the dual four-dimensional
field theory is formulated,
being $\R^4$, it is possible to have various types of infrared instabilities
if relevant perturbations are added.  In the AdS description, this
is related to the fact that the bad behavior of the metric at the
dangerous points $\vec y=\vec d_1,\ \vec d_2$ 
may be unstable against small perturbations.  
To compute correlation functions in these symmetry-breaking vacua,
care will be needed in specifying the desired behavior of perturbations
near the bad points of the metric; the crucial clue is presumably that
when there is only one such bad point, we know (via compactification of
the boundary) how to treat it.  In any event, we know from
the case of just one bad point that a ``singularity'' of this type
signals flow in the infrared to a non-trivial renormalization group
fixed point, the ${\cal N}=4$ Yang-Mills theory with gauge groups
determined by the coefficients of the singularities in $H$.

\newsec{Threebranes on the Conifold}

In this section, we will consider symmetry breaking in a more
complex example -- threebranes near the conifold singularity.
We recall that the conifold can be described in terms of
complex variables $w_a$, $a=1,\dots,4$ by the equation
\eqn\juxox{\sum_aw_a^2=0.}
Alternatively, one can introduce a $2\times 2$ matrix of complex
variables $m_{ij}$, $i,j=1,2$, and write
\eqn\ubox{0=\det \,m=m_{11}m_{22}-m_{12}m_{21}.}
The two descriptions are related by an obvious linear change of
variables.
We denote the manifold described by either of these equations
as $Y_6$; it is a cone over a five-manifold $X_5$, also denoted
as $T^{1,1}$.  $T^{1,1}$ is a homogeneous space $(SU(2)\times SU(2))/U(1)$,
where the $U(1)$ is a diagonal subgroup of $SU(2)\times SU(2)$ \Romans.

\def\N{{\bf N}}
\def\bar{\overline}
The near horizon geometry of a system of $N$ Type IIB threebranes at the 
conifold singularity is $AdS_5\times T^{1,1}$.  In \KW, we considered
a conformal field theory that is dual to Type IIB on that spacetime.
In this conformal field theory, the gauge group is $SU(N)\times SU(N)$
and there are chiral superfields $A_i$, $B_j$, $i,j=1,2$, with
$A_i$ transforming
as $(\N,\bar \N)$ and $B_j$ transforming as $(\bar\N,\N)$ under
$SU(N)\times SU(N)$.  There is also a superpotential $W=\epsilon^{ij}
\epsilon^{kl}\Tr A_iB_kA_jB_l$.

In \KW, we considered only the vacuum of the gauge theory with zero
expectation value for $A_i$ and $B_j$, and only the $AdS_5\times T^{1,1}$
solution on the string theory side.  Our intent here is to consider
more general vacua and solutions.

On the supergravity/string theory side, we can, to begin with,
consider the following three operations:

1) Deformation of the singularity.

2) Resolution of the singularity.

3) Moving the branes away from the singularity.

We will want to interpret all of these operations -- to the extent
that they can occur -- in tems of Higgsing of the gauge theory.

Deformation of the singularity means merely that an additional term
is added to the equation describing the conifold.  It becomes
\eqn\hdc{\sum_aw_a^2=\epsilon.}

Resolution of the singularity is more subtle to describe.\foot{The
resolution of interest to us is a ``small resolution'' that preserves
the Calabi-Yau condition.  If one did not wish to preserve the Calabi-Yau
condition, many more general resolutions would be possible.}  One
``solves'' the equation \ubox\ by introducing complex variables
$a_i$, $b_j$, $i,j=1,2$, and writing $m_{ij}=a_ib_j$.  Then one
imposes  the constraint
\eqn\hutullla{\sum_i|a_i|^2-\sum_j|b_j|^2= \delta,}
with $\delta$ a constant.  If $\delta=0$, then one gets a description of the
conifold by imposing also the equivalence relation
\eqn\tullgo{a_i\to e^{i\theta}a_i,~~b_j\to e^{-i\theta}b_j,}
to remove the redundancy in the ``solution'' $m_{ij}=a_ib_j$.
The resolution of the conifold is described by taking $\delta\not=0$,
and still imposing the equivalence relation.

$AdS_5\times T^{1,1}$ is the same topologically as $\R^4\times \R\times 
T^{1,1}$,
where $\R\times T^{1,1}$ is the cone $Y_6$ over $T^{1,1}$ with the singularity 
of the cone omitted. The usual metric on $AdS_5\times T^{1,1}$ is
\eqn\kilok{ds^2=H^{-1/2}\sum_i dx_i^2 +
H^{1/2}\left(dr^2+ r^2d\Omega^2 \right ),}
where $H$ is the standard Green's function $H=L^4/r^4$, $d\Omega^2$
is the Einstein metric \co\ on $T^{1,1}$, and $\sum_idx_i^2$
is the standard flat metric on $\R^4$.  In \kilok, $dr^2+r^2d\Omega^2$ should
be interpreted as a Calabi-Yau metric on $\R\times T^{1,1}$.
In the spirit of the present paper, we will incorporate symmetry
breaking by replacing $H$ by a more general Green's function
to account for motion of the threebranes, and by replacing
$dr^2+r^2 d\Omega^2$ by a Calabi-Yau metric on the deformation
$Y_6'$ of the conifold, or its resolution $Y_6''$.  (These Calabi-Yau
metrics have been discussed in \cd.)
  Our goal will then be to match this
more general geometry to a symmetry breaking vacuum in the gauge theory.

The two operations of deforming and resolving  the conifold are 
usually described in string theory in terms of motion in complex
structure or K\"ahler moduli space.  In the case of the conifold, these
motions are mutually
incompatible;  one can do either one but not both.
(There can be under suitable conditions a phase transition
between the two  branches \GMS.)
In the present context, only the resolution of the conifold is
generically allowed, in the following sense.  The manifold $T^{1,1}$ has
second Betti number 1, and as a result, in string theory on
$AdS_5\times T^{1,1}$ there are RR and NS-NS $B$-fields.
These combine together into a complex parameter which is interpreted
in terms of the gauge couplings of the $SU(N)\times SU(N)$ theory.
On the other hand, the deformation $Y_6'$ of the conifold
is topologically $T^*S^3$ (the cotangent bundle of the three-sphere)
and in particular its second Betti number vanishes.  Hence, in string
theory on $\R^4\times Y_6'$, any flat $B$-field can be gauged away.
If there are nonzero theta angles at infinity (that is, on $\R^4\times T^{1,1}$
-- we recall that $Y_6'$ looks like $T^{1,1}$ at infinity) --
then there must be non-zero curvatures $H=dB$ in the interior of $Y_6'$.
Non-zero values of the $H$-fields break supersymmetry, so one would
not get supersymmetric vacua in this way.  Moreover, the nonzero $H$ would
provide a source for the dilaton, so the fields with non-trivial
$B$-fields at infinity may not give $\R^4\times Y_6'$ solutions at all.
We conclude, then, that the deformation of the conifold is possible
(or at least, accessible) only if the $B$-fields vanish at infinity,
and thus only for special values of the gauge couplings of the dual
conformal field theory.  A phenomenon that can occur only for special
values of the gauge couplings is beyond our present understanding of the
gauge theory, and thus we will not attempt to further
analyze the deformation of the conifold in the present paper.
Furthermore, this particular case may turn out to be the hardest
to analyze because for $B=0$ one of the two $SU(N)$
gauge couplings is expected
to blow up \MP. This is natural from 
the point of view of flowing to the conifold field theory from
an orbifold field theory. Also, in the ``T-dual'' description in
terms of the NS5-branes and D4-branes, this is the point in the moduli
space where the distance between the two NS5-branes vanishes \Ur,
presumably giving rise to tensionless strings.
So the locus for which deformation of the conifold is possible is probably
quite subtle to describe.

On the other hand, the resolution $Y_6''$ is topologically an $\R^4$ bundle
over $S^2$; its second Betti number is 1.  Thus, the flat $B$-fields
on $T^{1,1}$ extend over $Y_6''$, and hence should make sense for
generic values of the gauge couplings.  Consequently, in the rest of this
section we concentrate on matching the resolution of the conifold
and the motion of the branes with phenomena in the gauge theory.

\subsec{ First Look At The Gauge Theory}

For simplicity, we will look at vacua in which $SU(N)\times SU(N)$ is
broken to a diagonal $SU(N)$.  Once these vacua are understood, the 
extension to more general cases is apparent.

To have an unbroken diagonal $SU(N)$, the $N\times N$
matrices $A_i$ and $B_j$ of chiral superfields must be (perhaps
after a gauge transformation) multiples of the identity.  So to get
vacua of this type, we set $A_i=a_i$, $B_j=b_j$, with complex numbers
$a_i$ and $b_j$.  The order parameters $m_{ij}=\langle\Tr\,A_iB_j\rangle/N$ 
are thus equal
to $a_ib_j$.  As we have already mentioned, equation \ubox\ is an immediate
consequence of $m_{ij}=a_ib_j$.  On the other hand, there is no
restriction in the gauge theory on the value of
$\delta =\sum_i|a_i|^2-\sum_j|b_j|^2$.
The gauge theory has both vacua with $\delta=0$ and vacua with 
$\delta\not= 0$.
We interpret this to mean that the dynamics of the gauge theory
includes a description of the resolution of the conifold.
The gauge theory also lacks the equivalence relation \tullgo.
We interpret this to mean that a pseudoscalar mode in $\R^4\times Y_6''$
that is related by supersymmetry to the resolution of the conifold
is likewise described by the gauge theory dynamics.  
The mode in question can be described as follows: it is a mode
of the four-form potential that transforms as a two-form on $\R^4$ times
a two-form on $Y_6''$.  We recall that a two-form on $\R^4$ is dual
to a scalar. 

In short, we propose that the resolution (but of course, not the deformation)
of the conifold is described by the dynamics of the dual $SU(N)\times SU(N)$
gauge theory.  According to this proposal,
the  resolution of the conifold is described
by the choice of vacuum in a fixed gauge theory with fixed coupling
constants, rather than by a change in the coupling constants
of the theory.  

Here is a bit of topological evidence for this proposal.
The $SU(N)\times SU(N)$ gauge theory with the chiral superfields
$A_i$, $B_j$, has a ``baryon number'' global symmetry
$A_i\to e^{i\theta}A_i$, $B_j\to e^{-i\theta}B_j$.  This symmetry
might be called ``baryon number'' because typical order parameters
are the baryonic, or dibaryonic, operators $\det\,A_i$ and $\det\,B_j$.
According to \GKtwo, the baryon number is mapped in the dual
$AdS_5\times T^{1,1}$ description to the wrapping number of threebranes
on a three-cycle in $T^{1,1}$.  Such a wrapping number exists because
the third Betti number of $T^{1,1}$ is 1.  (Topologically $T^{1,1}$ is
$S^3\times S^2$ \cd.)  Now, when we replace $AdS_5\times T^{1,1}$
by $\R^4\times Y_6''$, this conserved threebrane wrapping number no longer
exists, because the third Betti number of $Y_6''$ is zero.
(A wrapped threebrane at infinity
in $\R^4\times Y_6''$ can move into the interior
and annihilate.)  Thus, $\R^4\times Y_6''$ must correspond in the $SU(N)\times
SU(N)$  gauge
theory to vacua in which the baryonic charge is not conserved.
Indeed, in vacua with $\delta\not= 0$, the $a_i$ and $b_j$ cannot all
vanish, and hence the expectation values of
operators $\det\,A_i$ and $\det\,B_j$ carrying
baryon number are likewise not all zero.

At the cost of jumping slightly ahead of our story, we can also consider
the case that the gauge symmetry is broken, with $m_{ij}\not= 0$,
but $\delta =0$ and the conifold singularity is not resolved.
Also in this case, some of the baryonic order parameters $\det\,A_i$
and $\det \,B_j$ of the field theory are nonzero, so baryon number should
not be conserved.  On the AdS side, we will interpret these vacua
in terms of string theory on $\R^4 \times Y_6$, with a Green's function
$H$ whose singularity is not at the conical singularity of $Y_6$.
It follows that in this case, the conifold singularity is at a finite
distance in spacetime, and a wrapped threebrane can presumably disappear 
by collapsing to the conifold singularity.  This contrasts with
the $AdS_5\times T^{1,1}$ spacetime, which is dual to a vacuum with
unbroken symmetries; here the conifold singularity has disappeared
``to infinity'' in spacetime, and there is no way for a wrapped threebrane
to annihilate.

\subsec{Quantitative Treatment}

We now wish to describe these vacua somewhat more quantitatively.
The first step is to find the appropriate Calabi-Yau metric $ds^2_6$
on $Y_6$ or $Y_6''$.  Then one finds the appropriate Green's function
$H$ on $Y_6$ or $Y_6''$, with sources at the desired
positions of the threebranes, and the spacetime metric takes the familiar
form:
\eqn\cujku{ds^2=H^{-1/2}\sum_idx_i^2+H^{1/2}ds_6^2.}

This program can be described most explicitly for the case
of $Y_6$ -- moving the threebranes away from the conifold singularity
without resolving it -- because the Calabi-Yau metric on $Y_6$ is
particularly elementary.  In terms of the description of the conifold
via an equation $\sum_{a=1}^4 w_a^2=0$, set
\eqn\dolfo{\sum_{a=1}^4\bar w_aw_a=\rho^2.}
Then the $AdS_5$ radial coordinate is $r=\sqrt{3/2} \rho^{2/3}$.
The $Y_6$        metric has the familiar conical form
$ds^2_{\rm cone} =dr^2+r^2ds_5^2$.  The Laplace
equation for the Green's function away from the source is 
\eqn\tomigo{-{1\over r^5}{\partial\over\partial r}\left(r^5{\partial H
\over \partial r}\right)+{E\over r^2}H=0,}
where $E$ is the angular Laplacian on $T^{1,1}$.

Because $T^{1,1}$ is a homogeneous space, the spectrum of the angular
Laplacian can be worked out via group theory \refs{\SG,\ceresole}.  
$T^{1,1}$ has symmetry
group $SU(2)\times SU(2)\times U(1)$, where the $w_a$ transform
as $(1/2,1/2)$ under $SU(2)\times SU(2)$ and with charge 1 under
$U(1)$.  (The $U(1)$ is an R-symmetry group.)  The spherical
harmonics that are relevant for studying expectation values of chiral
superfields are the modes that transform as $(k/2,k/2)$ with $U(1)$
charge $\pm k$.  The corresponding wavefunctions are simply
\eqn\bub{\widehat w_{a_1a_2\dots a_k}={
w_{a_1}w_{a_2}\dots w_{a_k}\over |\sum_b\bar w_bw_b|^{k/2}},}
or the complex conjugate of this to reverse the sign of the $U(1)$
charge.  (The reason that these are the relevant modes is roughly
that modes with both $w$'s and $\bar w$'s in the numerator would
have a larger eigenvalue of the Laplacian for given $U(1)$ charge
and would ultimately lead to nonchiral operators.)
The eigenvalue of the Laplacian for these modes is \refs{\SG,\ceresole}
\eqn\honeypot{E(k)=3\left(k(k+2)
-{k^2\over 4}\right).}

If we look for a contribution to $H$ of the form $r^{c(k)}\widehat w_{a_1a_2
\dots a_k}$, we find that the equation \tomigo\ implies
\eqn\turnip{c(k)(c(k)+4)= E(k).}
We want the negative root, since we want $H$ to behave as $r^{-4}$ at
infinity.  So
\eqn\jucno{c(k)=-2-\sqrt{E(k)+4}.}
With the given values of $E(k)$, this gives the attractively simple result
\eqn\gunbo{c(k) =-4-3k/2.}
Given this, it follows that the relevant terms in $H$ take the form
\eqn\termbo{H={L^4\over r^4}\left(1+\sum_{k=1}^\infty{\left(f_{a_1\dots a_k}
\widehat w_{a_1\dots a_k}+c.c.\right)\over r^{3k/2}}\right)}
for some $f$'s.

The series of correction terms to $H$ of relative order $r^{-3k/2}$
give corrections to the $AdS_5\times T^{1,1}$ metric that vanish
like $r^{-3k/2}$ or $z^{3k/2}$ near the boundary.  According to 
eqn. \expect, such corrections imply that an operator of dimension
$3k/2$ has an expectation value.  For the harmonics described
above, the $U(1)$ or $R$-charge is $k$.  A field of $R$-charge $k$
and dimension $3k/2$ is a chiral superfield.  In this case,
the chiral superfields in question transform like $(k/2,k/2)$
under $SU(2)\times SU(2)$.  These are the quantum numbers of
the chiral superfields discussed in \KW, namely
$\Tr\,A_{i_1}B_{j_1}A_{i_2}B_{j_2}\dots
A_{i_k}B_{j_k}+{\rm permutations}$ of indices $i_1,\dots,i_k$
and $j_1,\dots,j_k$.  

Notice that for $k=1$, we have here an expectation value of an operator
of dimension $3/2$, which is above the unitarity bound (which is 1 in 
$d=4$ dimensions), but below the naive AdS bound of $d/2=2$.  The
formalism for studying in AdS space a scalar operator of dimension
in this range was discussed in section 2.  For our present purposes,
the net effect is simply that the term of relative order
$r^{-3k/2}$ in the metric is due to the expectation value
of a dimension $3k/2$ operator both for $k=1$ and for $k>1$.

So far, we have made no assumption about the nature of the source
terms for $H$: we have merely assumed that near $r=\infty$, $H$ obeys
the Laplace equation and vanishes as $r^{-4}$.  We can, if we wish,
require further that all threebranes are located at a point
$w_a=\epsilon_a$ on $Y_6$ away from the conifold singularity.
  In this case, $H$  will be invariant
under the subgroup of $SU(2)\times SU(2)\times U(1)$ that leaves
 fixed the point $w_a=\epsilon_a$.  
This implies that $f_{a_1a_2\dots a_k}$ in \termbo\ is a multiple of
$\bar\epsilon_{a_1}\bar\epsilon_{a_2}\dots \bar\epsilon_{a_k}$.
In particular, for $k=1$, we have $\sum_af_a^2=0$, and $f$ is
subject to no other restriction.  Translated into the language
of the gauge theory, this amounts to the statement that for a vacuum
with an unbroken diagonal $SU(N)$,
the order parameters $m_{ij}=\langle \Tr\, A_iB_j\rangle $ are coordinates
of a point on $Y_6$.  This is the familiar statement that the equation
\ubox\ of the conifold is a consequence of $m_{ij}=a_ib_j$.  Moreover,
 as explained in \KW,  for these Higgsed vacua with unbroken
diagonal $SU(N)$, the low energy theory has ${\cal N}=4$ supersymmetry
(even though the microscopic theory has ${\cal N}=1$).  This is the
result expected on the gravitational side, since with $\epsilon_a\not= 0$
the metric singularity is that of $N$ threebranes at a smooth point.

One can make a similar comparison between gauge theory and gravity
for the expectation values of chiral superfields
with $k>1$.  
For single trace operators,
which correspond to linear terms in the expansion of
$H^{1/2}$, these additional comparisons, to the extent that they can
be made without further detailed 
computations, yield little that is really
new since the results are nearly determined by the symmetries.  Both the 
expectation values of the chiral fields and the coefficients in the expansion
of $H$ are determined, up to a $k$-dependent 
normalization constant that depends on quantities we have not computed
(such as the precise constants in \termbo\ if $H$ has only a single
delta function source term at $w_i=\epsilon_i$),
by the unbroken symmetries. However, similarly to the example of section
3, for $k>1$ one finds that the coefficient of
the $r^{-3k/2}$ term in the metric is corrected by
non-linear terms in the expansion of $H^{1/2}$. Their structure
once again seems to correspond to multiple-trace operators in the
gauge theory, and the meaning of this requires further investigation.

\bigskip\noindent{\it The Resolution}

Threebranes on the resolved manifold $Y_6''$ can be described
similarly, the main difference being that the Calabi-Yau metric
of $Y_6''$ is known less explicitly \cd, and the description of the
Green's function $H$ will be correspondingly less explicit.

The resolution of the conifold can be interpreted, in gauge theory
language, in terms of giving an expectation value to a certain
operator ${\cal U}$.
Let us compute the dimension of this operator.  Under scalings of
$r$, the conical metric $dr^2+r^2d\Omega^2$ on $Y_6$ scales, obviously,
like $r^2$, as therefore does
the K\"ahler form $\omega$ on $Y_6$.  The resolution of $Y_6$
is obtained by a motion in K\"ahler
moduli space, and so by a topologically non-trivial correction to $\omega$.
This topologically nontrivial correction is indeed
\eqn\tiko{ \omega'={\epsilon^{abcd}w_a\overline w_bdw_cd\overline w_d
\over \left|\sum_a\overline w_a w_a\right|^2}.}
(The sign with which $ \omega'$ is added to the K\"ahler form determines
which of two topologically distinct resolutions of $Y_6$ is obtained.)
To verify that this is the correct $\omega'$, we proceed as follows.
$\omega'$ is invariant under
the $SO(4)$ symmetry of $Y_6$ but
not under the disconnected component of $O(4)$.  In fact, the disconnected
component of $O(4)$ exchanges the $a_i$ and $b_j$ in \hutullla, and
hence changes the sign of $\delta$ and therefore of $\omega'$.
Direct computation shows
that $d\omega'=0$ (this fixes the power of $|\overline w w|$ in the
denominator of \tiko\ and hence the scaling weight of $\omega'$).
However, $\omega'$ is not of the form $d\lambda$
for any $\lambda$.  (Indeed, by averaging over the compact group $SO(4)$, one
could always assume that $\lambda$ is $SO(4)$-invariant; but on $Y_6$
every $SO(4)$-invariant one-form is also $O(4)$-invariant.)

{}From \tiko, we see that $\omega'$ is invariant under scalings
of $w$ or equivalently of $r$, so it scales like $r^{-2}$ relative
to the unperturbed K\"ahler form $\omega$ on the cone.  This scaling
corresponds to the expectation value of an operator of dimension two
in the gauge theory.  
The term of order $r^{-2}$ 
is the leading
large $r$ correction in the C-Y metric on $Y_6''$:
the Einstein equations force
the topologically trivial corrections to $\omega$ to vanish faster than
$r^{-2}$ for large $r$.
There are also additional corrections to the
metric of $AdS_5\times T^{1,1}$ coming from the Green's
function $H$, similar
to the ones that were present in the symmetry-breaking vacua on $Y_6$;
these terms correspond to expectation values of the chiral operators
$\Tr (AB)^k$.

Let us try to interpret the correction 
term of order $r^{-2}$ in the gauge theory.
Comparing again to \hutullla, the natural gauge theory order parameter for
the resolution, in terms of the chiral superfields $A_i$ and $B_j$,
 is ${\cal U}=\Tr \,A_i\overline A_i- \Tr\,B_j\overline
B_j$.   We note that classically ${\cal U}$ has dimension two, in
agreement with the dimension that we found from the asymptotic
correction to the K\"ahler metric on $Y_6''$.  The operator ${\cal U}$
is contained in the same multiplet with the current that generates
the ``baryon number'' symmetry $A_i\to e^{i\theta}A_i$, $B_j\to
e^{-i\theta}B_j$.  The conserved current has no anomalous dimension,
so likewise the dimension of ${\cal U}$
is uncorrected in going from the classical description to supergravity.

Conserved current multiplets are among the several possible shortened
multiplets of $SU(2,2|1)$ \refs{\FGPW,\ceresole}.
As explained above, the operator ${\cal U}$ belongs to a 
conserved current multiplet and it is interesting to ask what is the
supergravity multiplet related to it through the AdS/CFT correspondence.
The multiplets appearing in type IIB supergravity 
on $AdS_5\times T^{1,1}$ were recently
classified in \ceresole, and we will make use of this analysis.
We expect the baryon number current of the gauge theory
to correspond to the massless gauge field in $AdS_5$ which couples
to the D3-brane wrapped over the 3-cycle in $T^{1,1}$.
This field is the component of the 4-form with one $AdS_5$ index and
three $T^{1,1}$ indices, $A_{\mu abc}$. Vector multiplet I
listed in Table 7 of 
\ceresole\ contains precisely this kind of vector field.
When the internal Laplacian eigenvalue is $E=0$, corresponding to
a singlet under $SU(2)\times SU(2)\times U(1)_R$, then this vector
field is massless. In this case the vector multiplet
in Table 7 also contains a scalar
in $AdS_5$ with $m^2=-4$ corresponding to dimension $\Delta=2$,
and we identify this field with the scalar operator
${\cal U}$. This field is a graviton with two $T^{1,1}$ indices
\ceresole, as expected from the preceding discussion.
To summarize,
the operator ${\cal U}$ and the baryon number current are related
through the AdS/CFT duality to fields from a massless $AdS_5$
vector multiplet.

\bigskip
\appendix{A}{Chiral Primary Operators}

Here we discuss the supergravity modes
which correspond to chiral primary operators. (For an extensive
analysis of the spectrum of the model that appeared following the original
version of the present paper, see \ceresole.) This will give
further background for the discussion in sections 3 and 4.
For the $AdS_5\times S_5$
case, these modes are mixtures of the conformal factors of the
$AdS_5$ and $S_5$ and the 4-form field. 
The same has been shown to be true for the
$T^{1,1}$ case \refs{\SG,\RD,\ceresole}. 
In fact, we may keep the discussion of such modes quite
general and consider $AdS_5\times X_5$ where $X_5$ is any Einstein manifold.

The diagonalization of such modes carried out by Kim, Romans and
van Nieuwenhuizen for the
$S^5$ case \Kim\ is easily generalized to any $X_5$.
The mixing of the conformal factor and 4-form modes results in
the following mass-squared matrix,
$$ m^2 = \pmatrix{ E+32 &  8E\cr  4/5 & E\cr}
$$
where $E\geq 0$ is the eigenvalue of the Laplacian on $X_5$.
The eigenvalues of this matrix are
\eqn\masses{ m^2 = 16 + E \pm 8 \sqrt{ 4+E}
\ .
}
We will be primarily interested in the modes which correspond
to picking the minus branch: they turn out to be the chiral primary
fields. For such modes there is a possibility of $m^2$ falling in
the range \range\ where there is a two-fold ambiguity in defining
the corresponding operator dimension. This happens for the eigenvalue $E$
such that
\eqn\bound{5 \leq E \leq 21 \ .
}

First, let us recall the $S^5$ case where the spherical
harmonics correspond to
traceless symmetric tensors of $SO(6)$, $d^{(k)}_{i_1\ldots i_k}$. 
Here $E= k(k+4)$, and it seems that
the bound \bound\ is satisfied for $k=1$. However, this is precisely the
special case where the corresponding mode is missing.
For $k=0$ there is no 4-form mode, hence no
mixing, while for $k=1$ one of the mixtures is the singleton \Kim.
Thus, all chiral primary 
operators in the ${\cal N}=4$ $SU(N)$ theory correspond to
the conventional branch of dimension, $\Delta_+$.
It is now well-known that this family of operators with dimensions
$\Delta= k$, $k=2,3,\ldots$
is $d^{(k)}_{i_1\ldots i_k} \Tr (X^{i_1} \ldots X^{i_k})$.
The absence of $k=1$ is related to the gauge group being $SU(N)$ rather
than $U(N)$. Thus, in this case we do not encounter operator
dimensions lower than $2$.

The situation is different for $T^{1,1}$. Here there is a family of wave
functions labeled by non-negative integer $k$, transforming under 
$SU(2)\times SU(2)$  as $(k/2,k/2)$, and with
$U(1)$ charge $k$.
They are described in section 4.2, and the eigenvalues of the Laplacian
are written in \honeypot.
In \KW\ it was argued that the corresponding chiral operators are
$$\Tr (A_{i_1} B_{j_1} \ldots A_{i_k} B_{j_k} )
\ .
$$
Since the F-term constraints in the gauge theory require that the
$i$ and the $j$ indices are separately symmetrized, we find that
their $SU(2)\times SU(2)\times U(1)$ quantum numbers agree with those
given by the supergravity analysis. Since in the field theory construction
of \KW\ the $A$'s and the $B$'s have dimension $3/4$, the dimensions
of the chiral operators are $3k/2$.

In studying the dimensions from the supergravity point of view, one encounters
a subtlety discussed in section 2. While for $k>1$ only
the dimension $\Delta_+$ is admissible, for $k=1$ one could pick either 
branch. Indeed, from \honeypot\ we have
$E(1)=33/4$ which falls within the range \bound. Here we
find that $\Delta_-=3/2$, while $\Delta_+=5/2$. Since the supersymmetry 
requires the corresponding dimension to be $3/2$, in this case
we have to pick the unconventional $\Delta_-$ branch. Choosing this
branch for $k=1$ and $\Delta_+$ for $k>1$ we indeed find
following \SG\ 
that the supergravity analysis based on \branches, \masses, \honeypot,
reproduces the dimensions $3k/2$.
Thus, the conifold theory provides
a simple example of the AdS/CFT duality where the $\Delta_-$ branch
has to be chosen for certain operators.

Let us also note that substituting $E(1)=33/4$ into \masses\ we find
$m^2=-15/4$ which corresponds to a conformally coupled scalar in $AdS_5$
\Kim. In fact, the supermultiplet containing this scalar has to include
another conformally coupled scalar and a massless fermion. One of these
scalar fields corresponds to the lower component of the superfield
$\Tr (A_i B_j)$, which has dimension $3/2$, while the other
corresponds to the upper component which has dimension $5/2$. Thus,
the supersymmetry requires that we pick dimension $\Delta_+$ for one
of the conformally coupled scalars, and $\Delta_-$ for the other.

\bigskip
\noindent
{\bf Acknowledgements}
We are grateful to A. Ceresole, G. Dall'Agata, D. Freedman,
S. Gubser, P. Kraus, F. Larsen, V. Periwal and
L. Rastelli for useful comments.
The work  of I.R.K. was supported in part by the NSF
grant PHY-9802484 and
by the James S. McDonnell
Foundation Grant No. 91-48.
The work of E.W. is supported in part by NSF grant PHY-9513835.
\vfill\eject
\listrefs
\end